# *K*-Slicing the Reissner-Nordstrom Spacetime: Some New Observations

by

Asghar Qadir[1], M. Sajid[2] and Azad A. Siddiqui[3,4]

**Abstract**

There were problems encountered in extending the *K*-slicing of the Schwarzschild and Reissner-Nordstrom (RN) spacetimes [1, 2] to the extreme case, when charge equals mass (in gravitational units). The earlier procedure is here modified so as to allow us to obtain a *K*-slicing of the region outside the horizon of the extreme metric by spacelike hypersurfaces. We checked this new procedure by applying it to the Schwarzschild and usual RN metrics and recovering the previous foliation with an improved accuracy. We have also checked the asymptotic behaviour of the *K*-slicing for large *K* by extrapolation.

## 1. Introduction

The Schwarzschild spacetime in Schwarzschild coordinates is given by

$$ds^2 = (1 - r_s/r)dt^2 - \frac{dr^2}{(1 - r_s/r)} - r^2 d\theta^2 - r^2 \sin^2\theta\, d\phi^2, \quad (1)$$

where $r_s$ is the Schwarzschild radius (2m in gravitational units). Brill et al. [3] use the

[1] Centre for Advanced Mathematics and Physics, Campus of EME College, National University of Science and Technology, Peshawar Road, Rawalpindi, Pakistan; and Department of Mathematical Sciences, King Fahd University of Petroleum and Minerals, Dhahran, 31261. E-mail: aqadirs@comsats.net.pk
[2] Department of Mathematics, Quaid-i-Azam University, Islamabad, Pakistan.
[3] Department of Basic Science &Humanities, EME College, National University of Science and Technology, Peshawar Road, Rawalpindi, Pakistan. E-mail: azad@ceme.edu.pk
[4] *Present Address:* Department of Physics and Measurement Technology, Linköping University, SE-583 81 Linköping, Sweden. E-mail: azad@ifm.liu.se

Kruskal-Szekres (KS) coordinates $(v, u)$ given by [4]

$$u = (r/r_s - 1)^{1/2} \exp(r/2r_s)\cosh(t/2r_s),$$
$$v = (r/r_s - 1)^{1/2} \exp(r/2r_s)\sinh(t/2r_s), \quad (2)$$

in the region where $r > r_s$. They obtain an equation for hypersurfaces of constant mean extrinsic curvature, $K$,

$$\frac{dv}{du} = \frac{Av + Eu}{Au + Ev}, \quad (3)$$

$$E = H - \frac{Kr^3}{3}, \qquad A = \sqrt{E^2 + r^3(r - r_s)}, \quad (4)$$

where $H$ is an arbitrary parameter that measures how much the intrinsic and extrinsic curvature vary on hypersurface. The foliation of spacetime by such hypersurfaces is called a $K$-slicing.

Eq. (3) was converted [1] to the compactified KS coordinates $(\psi, \xi)$ given by

$$\psi = \tan^{-1}(v - u) + \tan^{-1}(v - u)$$
$$\xi = \tan^{-1}(v - u) - \tan^{-1}(v + u), \quad (5)$$

which yields

$$\frac{d\psi}{d\xi} = \frac{A\sin\psi\cos\xi + E\sin\xi\cos\psi}{A\sin\xi\cos\psi + E\sin\psi\cos\xi}. \quad (6)$$

For the RN spacetime given by

$$ds^2 = \left(1 - \frac{2m}{r} + \frac{Q^2}{r^2}\right)dt^2 - \frac{dr^2}{\left(1 - \frac{2m}{r} + \frac{Q^2}{r^2}\right)} - r^2 d\theta^2 - r^2 \sin^2\theta\, d\phi^2, \quad (7)$$

where $m$ and $Q$ are the mass and charge of the source (in gravitational units). Eq. (6) remains unchanged, with $A$ given by

$$A = \sqrt{E^2 + r^2(r - r_+)(r - r_-)}, \tag{8}$$

instead of eq. (4), where

$$r_\pm = m \pm \sqrt{m^2 - Q^2}. \tag{9}$$

The procedure adopted earlier to solve eq. (6) for foliating the Schwarzschild and RN spacetimes is the following [2]. Choose a particular value of $K$ and require that $\left.\frac{d\psi}{d\xi}\right|_{\xi=0} = 0$, which implies that $A = 0$ at $\xi = 0$ in eq. (6). This provides a relationship between $H$ and the initial value, $r_i$ of $r$:

$$H = \frac{Kr^3}{3} \pm \sqrt{r_i^3(r_s - r_i)}, \tag{10}$$

for the Schwarzschild spacetime and

$$H = \frac{Kr^3}{3} \pm r_i\sqrt{(r_+ - r_i)(r_- - r_i)}, \tag{11}$$

for the RN spacetime. Now, solve the equation numerically by choosing the initial value of the radial parameter, $r_i$, at the centre of the Carter-Penrose (CP) diagram for a particular value of $K$, and requiring that the hypersurfaces go to spacelike infinity, $I^0$, at $\xi = 0$. The calculation cannot be carried out up to $\xi = \pi$, as the expression become unstable. Consequently, carry out the calculation as far as possible with the requirement that the hypersurface should not hit either null infinity. Time constraints, and the onset of numerical instabilities, limit the maximum achievable $\xi$ for $\psi = 0$. It was found that one could go to higher values of $K$ for the RN than for the Schwarzschild spacetime.

For the extreme RN (eRN) spacetime given by

$$ds^2 = (1 - m/r)^2 dt^2 - \frac{dr^2}{(1 - m/r)^2} - r^2 d\theta^2 - r^2 \sin^2\theta \, d\phi^2, \qquad (12)$$

the central line in the CP diagram corresponds to $r = 4m$. Thus we cannot choose $r_i$ at the central line but, as will be explained shortly, can choose another coordinate on it. This was not the main problem faced. The real problem was the lack of stability with higher values of $Q/m$. This is tackled here by introducing a new procedure near $I^0$.

## 2. The New Procedure

The procedure adopted here, for the RN or Schwarzschild spacetimes, is that instead of solving eq. (6) as described above, we truncate it at $\xi = \pi - \varepsilon$, where $\varepsilon$ is a small number. We now require that the difference between the slope of the hypersurface given by eq. (6) and that of the calculated value at $\xi = \pi - \varepsilon$, be minimum. Before applying this new procedure to the eRN spacetime we applied it to the Schwarzschild spacetime and recovered the results obtained earlier with slightly improved accuracy. In the RN case the increase of accuracy is so great that we obtained the foliation for significantly higher values of $K$. By reducing $\varepsilon$ the inaccuracy due to linearizing the hypersurface at the end can be made arbitrarily small. In principle it is possible to use higher order terms in the Taylor series for the hypersurface but it is not clear that the accuracy would be adequately improved by it. For the Schwarzschild spacetime the improvement in accuracy is slight.

The results of the foliation of the usual RN spacetime are given in Table 1. There are problems of numerical instability when $Q/m$ is small. The hypersurfaces can be found for higher values of $K$ as $Q/m$ increases. We obtained the hypersurfaces for larger values of $K$ than the previous foliation. For example for $Q/m = 0.8$ previously the hypersurfaces went up to $K = 0.8$ and with the new procedure up to $K = 1.9$. The hypersurfaces for $Q/m = 0.6, 0.8$ and $0.95$ are shown in Fig. 1, 2 and 3 respectively.

### 3. Foliation of the eRN spacetime

Using the procedure of Brill et al. we obtained the $K$-slicing equation of the eRN spacetime in $(t, r)$ coordinates and is given by

$$\frac{dt}{dr} = \frac{r^2 E}{(r-m)^2 A}, \tag{13}$$

where $E$ is given by equation (2) and $A$ is modified to

$$A = \sqrt{E^2 + r^2(r-m)^2}. \tag{14}$$

There are no Kruskal or Kruskal-Szekres coordinates, or their analogues, for the eRN spacetime [5]. Consequently, we converted to Carter's $(\psi, \xi)$ coordinates [6] given by

$$\left.\begin{aligned}\psi &= \tan^{-1} v + \cot^{-1} w \\ \xi &= \tan^{-1} v - \cot^{-1} w\end{aligned}\right\}, \tag{15}$$

$$\left.\begin{aligned}v &= t + r^* \\ w &= -t + r^*\end{aligned}\right\}, \tag{16}$$

$$r^* = r - \frac{rm}{r-m} - 2m\ln\left|\frac{r-m}{m}\right|. \tag{17}$$

Here $-3\pi/2 < \psi < \pi/2$ and $-\pi/2 < \xi < 3\pi/2$ for the region outside the horizons $(\psi + \pi/2) = \pm(\xi - \pi/2)$ and $\psi = -\pi/2, \xi = 3\pi/2$ represents $I^0$, in the CP diagram.

In $(v, w)$ coordinates eq. (13) is given by

$$\frac{dv}{dw} = \frac{A+E}{A-E}, \qquad (18)$$

and in Carter's coordinates eq. (18) takes the form

$$\frac{d\psi}{d\xi} = \frac{A\cos\xi\cos\psi + E(1-\sin\xi\sin\psi)}{A(1-\sin\xi\sin\psi) + E\cos\xi\cos\psi}. \qquad (19)$$

Requiring that for the hypersurfaces $\left.\frac{d\psi}{d\xi}\right|_{\xi=\pi/2} = 0$ and they go to $I^0$ $(\psi = -\pi/2, \xi = 3\pi/2)$ which implies that $E = 0$ initially. Therefore $H = \frac{64Km^3}{3}$.

The results of the foliation of the eRN spacetime are given in Table 2 and the hypersurfaces are shown in Fig. 4. Notice that $K$ and $\psi$ go to very much higher values than for the Schwarzschild or RN cases. This shows that the utility of the new procedure increases consistently with increasing $Q/m$.

**4.     Extrapolation**

One would ideally like to carry on the numerical computation up to the domain where the analytical arguments of [2] can be applied. Due to time constraints of computation, we could not carry out the foliation to that extent. However, it will be noticed that the initial value of the relevant time parameter comes close to the limiting value by the end of the table. To carry the foliation further, the initial times must not exceed the limiting value. This can be checked by taking the initial time differences from one hypersurface to the next. In the beginning the differences change significantly. As such, we need to incorporate the second differences and even the third and fourth in some cases.

The results of the extrapolation of the RN spacetime are as follows: for $Q/m = 0.6$, $K = 254.5$, for $Q/m = 0.8$, $K = 213.9$, for $Q/m = 0.95$, $K = 118.2$ as shown in Fig. 5. These values are sufficiently high for the previous asymptotic analysis to be applied and hence it provides a complete foliation up to the inner horizon.

## 5. Conclusion

We have obtained a complete foliation of the interior of the eRN spacetime outside the horizon. This was achieved by improving the procedure used previously. There the "shooting method" was used to push the point where the hypersurface cuts the $\psi = 0$ line out to the maximum value of $\xi$. As $K$ increases the guesses become less reliable and numerical stability becomes a serious concern.

To validate the procedure used, the previous foliation of the Schwarzschild and RN spacetimes was first reproduced. For small values of $Q/m$ the procedure remains unstable but for larger values there was a significant improvement in the computing time and in the values of *K* for which the hypersurfaces could be computed. With this improved accuracy we were able to obtain the required foliation for the eRN metric. The behaviour of the extreme case consistently followed the behaviour of the RN foliations as *Q*/*m* increases.

The next step we took was to check whether the foliation is complete. This was done by extrapolating the value of the initial time, $\psi_i$, for larger *K* by using fourth difference where it could be used (and lower when necessary). The results showed consistent behaviour for sufficiently large *K*. It will be apparent from Fig. 5 that the extrapolation follows a nearly linear behaviour for larger *K*. This leads to the extrapolation of $\psi_i$ hitting $\pi$ for a finite *K*. Of course, the extrapolation would need to be asymptotic to $\psi_i = \pi$. This would be apparent if we could take higher differences into account. For this purpose we would need to compute over a finer scale of *K*, which would take too much computational time. Our purpose was already served when we found that we could compute the initial time for very large *K*.

It needs to be stressed out that our foliation is only of the *interior region* of the spacetime and not of the compactified spacetime. The attempt of Brill et al. was for the full compactified Schwarzschild spacetime. Thus they required that the hypersurfaces become asymptotically flattened out, while we took them to all go to $I^0$.

**Figure Captions**

Fig. 1: Foliation of the RN spacetime, by spacelike $K$-surfaces for $Q/m = 0.6$. The hypersurfaces are labeled according to their serial numbers in Table 1.

Fig. 2: Foliation of the RN spacetime, by spacelike $K$-surfaces for $Q/m = 0.8$. The hypersurfaces are labeled according to their serial numbers in Table 1.

Fig. 3: Foliation of the RN spacetime, by spacelike $K$-surfaces for $Q/m = 0.95$. The hypersurfaces are labeled according to their serial numbers in Table 1.

Fig. 4: Foliation of the eRN spacetime, by spacelike $K$-surfaces. The hypersurfaces are labeled according to their serial numbers in Table 2.

Fig. 5: Extrapolated values of $\psi_i$ and $K$ for the RN spacetime for $Q/m = 0.6$, $0.8$ and $0.95$.

**Table 1:** Initial Kruskal Szekres time, $\psi_i$, is given for different values of *K* and different values of *Q/m* for an RN spacetime *K*-surfaces For *Q/m* = 0.6 there are sixteen, for *Q/m* = 0.8 twenty-eight and for *Q/m* = 0.95 forty.

| No. | mK | $\Psi_i$ | | |
|---|---|---|---|---|
| | | *Q/m* = 0.6 | *Q/m* = 0.8 | *Q/m* = 0.95 |
| ±1 | 0.01 | ±0.3170 | ±0.48900000 | ±0.5470000 |
| ±2 | 0.02 | ±0.6080 | ±0.8850000 | ±1.0290000 |
| ±3 | 0.05 | ±1.2030 | ±1.63400000 | ±1.9160000 |
| ±4 | 0.10 | ±1.6160 | ±2.17800000 | ±2.4480000 |
| ±5 | 0.20 | ±1.7820 | ±2.38700000 | ±2.7080000 |
| ±6 | 0.30 | ±1.8278 | ±2.44100000 | ±2.7910000 |
| ±7 | 0.40 | ±1.8571 | ±2.46760000 | ±2.8290000 |
| ±8 | 0.50 | ±1.8713 | ±2.48400000 | ±2.8510000 |
| ±9 | 1.00 | | ±2.53274000 | ±2.8967600 |
| ±10 | 1.50 | | ±2.56366600 | ±2.9197100 |
| ±11 | 1.60 | | ±2.56858950 | ±2.9233600 |
| ±12 | 1.70 | | ±2.57318985 | ±2.9267940 |
| ±13 | 1.80 | | ±2.57749454 | ±2.9300340 |
| ±14 | 1.90 | | ±2.58152908 | ±2.9330960 |
| ±15 | 2.00 | | | ±2.9359970 |
| ±16 | 2.50 | | | ±2.9485000 |
| ±17 | 3.00 | | | ±2.9584520 |
| ±18 | 3.50 | | | ±2.9665828 |
| ±19 | 3.60 | | | ±2.9680351 |
| ±20 | 3.70 | | | ±2.9694371 |

**Table 2:** The initial Carter time, $\psi_i$, added to $\pi/2$, is given as $\Psi_i$ for different values of $K$ in the eRN spacetime ($Q = m$). Twenty-two surfaces are so described. Notice that for $K$ beyond $0.5/m$ the values of $\Psi_i$ have nearly converged. Clearly, as $K \to \infty$, $\Psi_i \to \pi$.

| No. | $mK$ | $\Psi_i = \psi_i + \pi/2$ |
|---|---|---|
| ±1 | 0.01 | ±0.9683000 |
| ±2 | 0.02 | ±1.9390000 |
| ±3 | 0.05 | ±2.7540000 |
| ±4 | 0.10 | ±3.0825000 |
| ±5 | 0.20 | ±3.1156000 |
| ±6 | 0.50 | ±3.1306700 |
| ±7 | 1.00 | ±3.1308380 |
| ±8 | 1.20 | ±3.1308760 |
| ±9 | 1.50 | ±3.1309201 |
| ±10 | 2.00 | ±3.1309826 |
| ±11 | 2.20 | ±3.1310054 |

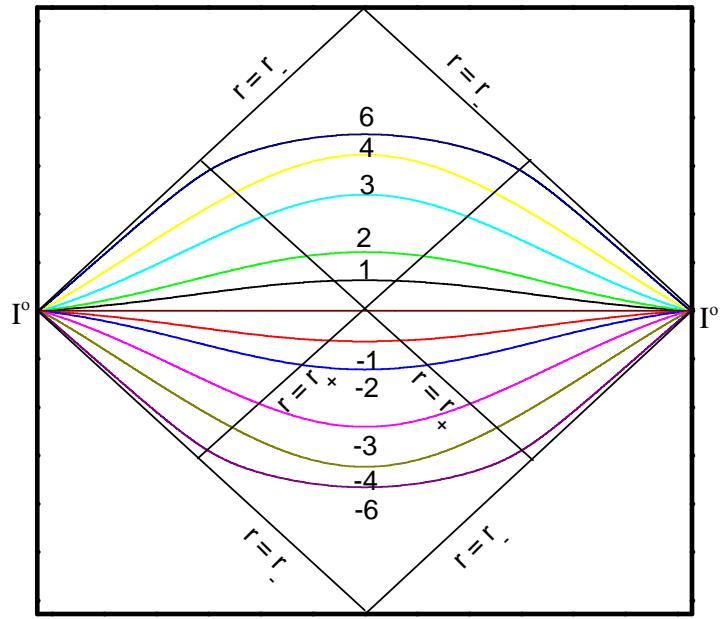

Fig. 1

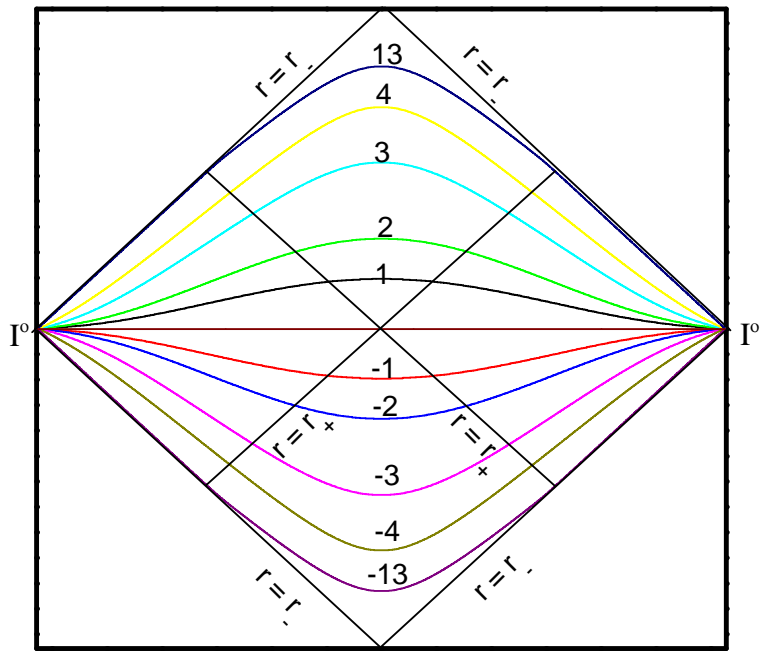

Fig. 2

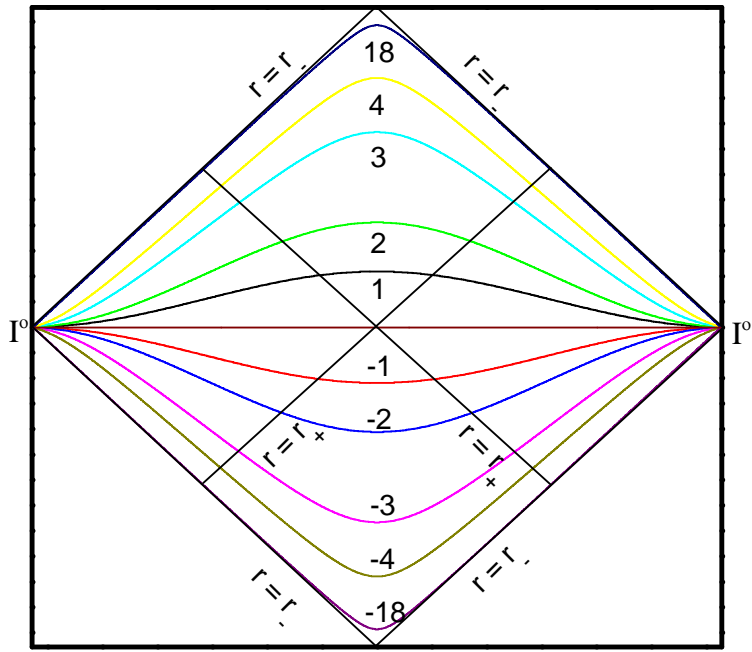

Fig. 3

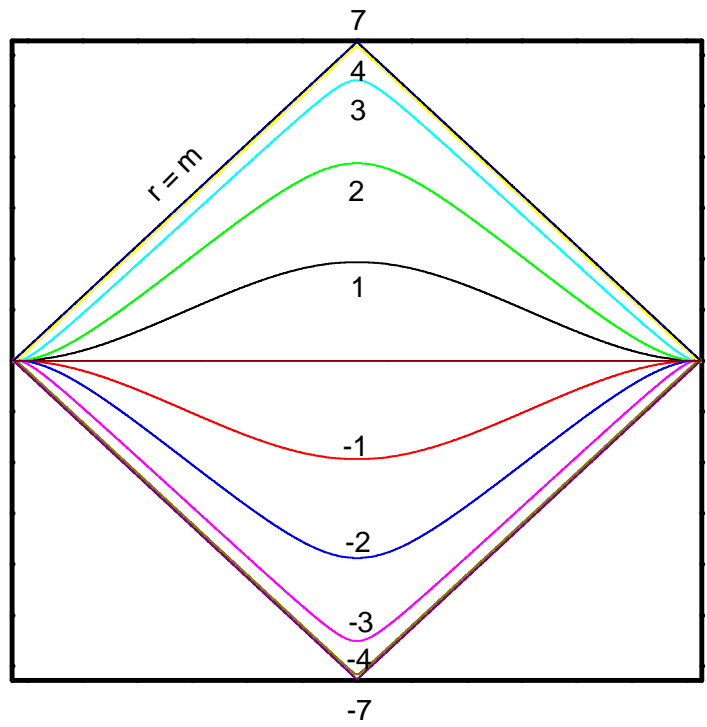

Fig. 4

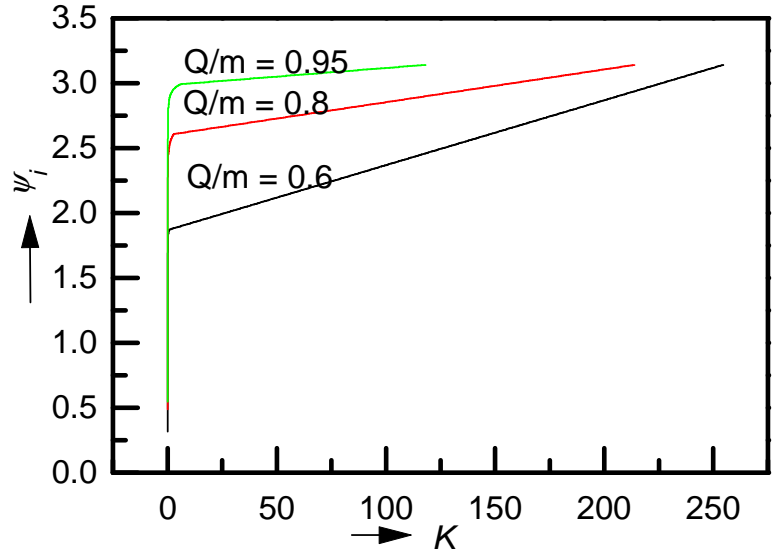

Fig. 5